\documentclass[aps,prl,twocolumn,showpacs,groupedaddress]{revtex4}
\usepackage{graphicx}

\begin{document}


\title{Two-fermion bound state in a Bose-Einstein condensate}
\author{Weiping Zhang, Han Pu, Chris P. Search, Pierre Meystre
and E. M. Wright}
\affiliation{Optical Sciences Center, The
University of Arizona, Tucson, AZ 85721}

\date{\today}

\begin{abstract}
A nonlinear Schr\"odinger equation is derived for the dynamics of
a beam of ultracold fermionic atoms traversing a Bose-Einstein
condensate. The condensate phonon modes are shown to provide a
nonlinear medium for the fermionic atoms. A two-fermion bound
state is predicted to arise, and the signature of the bound state
in a nonlinear atom optics experiment is discussed.
\end{abstract}

\maketitle

The recent experimental success in creating quantum-degenerate
atomic Fermi gases \cite{jin,hulet,salomon,ketterle,thomas} is
opening up fascinating new opportunities to explore the quantum
statistics of ultracold atoms. Of particular interest in this
context is the formation of atomic Cooper pairs in ultracold Fermi
gases \cite{molmer,tosi,stoof,smith,yip,pu,pair}. Another
important question is whether nonlinear atom optical effects,
which are now well established in bosonic systems, can also occur
with fermions. Fermionic behavior is strongly constrained by the
Pauli exclusion principle. This limits the variety of possible
nonlinear atom optics effects, but also offers the potential for
novel applications without analogs in optics. These include, for
example, low-noise inertial and rotation sensors, and quantum
information processing. Research on nonlinear atom optical effects
in a quantum-degenerate Fermi gas has recently been initiated. Two
theoretical papers \cite{MooMey01,KetIno01} have indicated that
fermionic atom four-wave mixing is possible under appropriate
conditions.

Our goal in this Letter is to explore a new situation where a gas
of bosons serves as a nonlinear medium for fermionic atoms. In
particular, we study how the interatomic interaction between a
Bose-Einstein condensate and a fermionic beam can be employed to
manipulate the quantum state of the beam. By drawing on the
analogy to nonlinear optics we describe the interaction in terms
of an effective attractive Kerr nonlinearity, and show that a
two-fermion bound state can result with a unique signature in a
nonlinear atom optical experiment.

Since the Pauli exclusion principle precludes the direct
evaporative cooling of spin-polarized fermionic samples, current
experiments employ either unpolarized fermionic mixtures
\cite{jin,thomas} or Bose-Fermi mixtures
\cite{hulet,salomon,ketterle}. Our starting point is a beam of
fermionic atoms with two internal spin states $|\uparrow \rangle$
and $|\downarrow \rangle$ interacting via two-body collisions with
a quantum degenerate Bose gas. The atomic Bose field is decomposed
in the familiar way as ${\hat \psi}_B({\bf r})= \phi_B({\bf r}) +
{\hat \xi}({\bf r})=\sqrt{n_B({\bf r})}+ {\hat \xi}({\bf r})$,
where $\phi_B({\bf r})$ is the condensate wave function, taken to
be real for simplicity, ${\hat \xi}({\bf r})$ describes the
elementary excitations above the condensate, and $n_B= |\phi_B|^2$
is the condensate density. The fermionic field is described by the
field operators ${\hat \psi}_\sigma({\bf r})$, with $\sigma =
\{\uparrow, \downarrow\}$. In the following, it will also be
useful to introduce the density fluctuations of the fermion fields
by $\delta{\hat n}_\sigma = {\hat \psi}^\dagger_\sigma({\bf
r}){\hat \psi}_\sigma({\bf r})- \langle{\hat
\psi}^\dagger_\sigma({\bf r}){\hat \psi}_\sigma({\bf r}) \rangle =
{\hat n}_\sigma - n_\sigma$.

The Bose and Fermi systems are coupled by two-body interactions.
In the shapeless approximation, and taking into account that
$s$-wave scattering is forbidden between fermionic atoms of same
spin, we find that to lowest-order in ${\hat \xi}({\bf r})$, the
dynamics of the bosonic atoms is given by the coupled equations
\cite{pu}
\begin{eqnarray}
\left(\hat{H}_B^{(0)}+gn_B({\bf r})+\sum_{\sigma} f_{\sigma}
n_{\sigma}({\bf r}) \right) \phi_B = \mu_B \phi_B, \nonumber \\
i\hbar \frac{\partial \hat{\xi}}{\partial t} =
\left(\hat{H}_B-\mu_B \right)\hat{\xi} + g\phi^2 \hat{\xi}^\dagger
+\phi \sum_{\sigma} f_{\sigma} \delta \hat{n}_{\sigma}. \label{b}
\end{eqnarray}
To the same order, the fermionic field equations are
\begin{eqnarray}
&&i\hbar \frac{\partial \hat{\psi}_{\sigma}}{\partial t} = \left(
\hat{H}_{F\sigma}-\mu_{\sigma} \right) \hat{\psi}_{\sigma}
\nonumber \\
&+& f_{\sigma}(\phi_B \hat{\xi}^\dagger + \phi_B^* \hat{\xi})
\hat{\psi}_{\sigma} + h \hat{\psi}^{\dagger}_{\sigma'}
\hat{\psi}_{\sigma'} \hat{\psi}_{\sigma}.\label{f}
\end{eqnarray}
Here, $\hat{H}_{\alpha}^{(0)}=\hat{T}_{\alpha} + V_{\alpha}
(\alpha = B, \sigma)$ are the single-particle Hamiltonians for
bosonic atoms and for fermionic atoms of spin $\sigma$,
respectively, $\hat{T}_\alpha$ and $V_\alpha$ are the associated
kinetic energy and trapping potential. The Hamiltonians
$\hat{H}_B=\hat{H}_B^{(0)}+2gn_B+\sum_{\sigma}
f_{\sigma}n_{\sigma}$, and $\hat{H}_{F
\sigma}=\hat{H}_{F\sigma}^{(0)}+f_{\sigma} n_B$, also include the
self-contribution to the mean-field energy of the respective
fields. Finally, $\mu_\alpha$ are the chemical potentials, and the
parameters $g$, $f$ and $h$ represent the boson-boson,
boson-fermion, and fermion-fermion interaction strengths. In terms
of the $s$-wave scattering lengths $a$, they are given by $g =
4\pi \hbar^2a_B/m_B,f_{\sigma} = 2\pi \hbar^2 a_{BF\sigma}/m_r,
h=4\pi \hbar^2a_{\uparrow \downarrow}/m_F$, with $m_r=m_F
m_B/(m_F+m_B)$ being the reduced mass.

In previous work \cite{pu} we discussed how the coupling to a
fermionic component can change the dynamical stability of a Bose
condensate. Here we concentrate instead on how the presence of a
condensate can induce nonlinear dynamics of a fermion field. To
isolate the key underlying physical mechanisms, it is useful to
simplify the situation as much as possible. With this in mind, we
consider a situation where the back-action of the fermionic fields
on the condensate is negligible. Specifically, we assume that $g
n_B >> \sum_{\sigma} f_{\sigma} n_{\sigma}$ is satisfied, in which
case we can ignore the effects of the fermionic beam on the
condensate wave function $\phi_B({\bf r})$. We then apply a
standard Bogoliubov approach\cite{fetter} to determine the effect
of the fermions on the excitation field ${\hat \xi}(\bf r)$. One
readily finds
\begin{widetext}
\begin{equation}
\hat{\xi}({\bf r},t) = \hat{\xi}^{(0)}({\bf r},t) +\frac{1}{i
\hbar} \int_0^t dt' \int d^3r' \left[ G({\bf r},{\bf r}',t-t')
\phi_B({\bf r}') - F({\bf r},{\bf r}',t-t') \phi^*_B({\bf r}')
\right] \sum_{\sigma} f_{\sigma} \delta {\hat n} _{\sigma}
\label{ph}.
\end{equation}
\end{widetext}
The first term $\hat{\xi}^{(0)}({\bf r},t)$ on the right-hand side
describes the free-field vacuum quasi-particle fluctuations in the
absence of fermions. It has the familiar form
\begin{eqnarray}
\hat{\xi}^{(0)}({\bf r},t)=\sum_{\bf n} \left( u_{\bf n}({\bf r})
e^{-iE_{\bf n} t/\hbar} \hat{\alpha}_{\bf n} - v^*_{\bf n}({\bf
r})e^{iE_{\bf n} t/\hbar} \hat{\alpha}^{\dagger}_{\bf n} \right),
\nonumber
\end{eqnarray}
with the Bogoliubov quasi-particle operators $\hat{\alpha}_{\bf
n}$ and $\hat{\alpha}^{\dagger}_{\bf n}$ satisfying Bose
commutation relations. The quasi-particle mode functions $u_n({\bf
r})$ and $v_n({\bf r})$, and corresponding energy eigenvalues are
determined by the matrix equations
\begin{equation}
\left[ \begin{array}{cc}
\hat{H}_B & -gn_B \\
gn_B & -\hat{H}_B
\end{array} \right]
\left[ \begin{array}{c}
u_{\bf n}({\bf r}) \\
v_{\bf n}({\bf r})
\end{array} \right]
= E_{\bf n} \left[ \begin{array}{c}
u_{\bf n}({\bf r}) \\
v_{\bf n}({\bf r})
\end{array}\right].
\end{equation}
The second term in Eq. (\ref{ph}) is a four-wave mixing process
that mixes the condensate with quasi-particles. It is mediated by
the density fluctuations of the Fermi fields, whose evolution is
governed by the quasi-particle Green's functions
\begin{eqnarray}
G({\bf r},{\bf r}',\tau)=\sum_{\bf n} \left( e^{\frac{-iE_{\bf n}
\tau}{\hbar}} u_{\bf n}({\bf r}) u^*_{\bf n}({\bf
r}')-e^{\frac{iE_{\bf n} \tau}{\hbar}} v^*_{\bf n}({\bf r}) v_{\bf
n}({\bf r}')\right) , \nonumber
\end{eqnarray}
with a similar form for $F({\bf r},{\bf r'},\tau)$, but with
$u^*_{\bf n}({\bf r}')$ and $ v_{\bf n}({\bf r}')$ replaced by
$v^*_{\bf n}({\bf r'})$ and $u_{\bf n}({\bf r'})$, respectively.

The lowest-order contribution of the condensate to the dynamics of
the Fermi fields is obtained by substituting Eq. (\ref{ph}) into
Eq. (\ref{f}). Using the homogeneous case quasi-particle mode
functions of Ref. \cite{fetter} for simplicity, this yields the
stochastic Heisenberg equations of motion
\begin{widetext}
\begin{equation}
i\hbar \frac{\partial \hat{\psi}_{\sigma}}{\partial t} = \left(
\hat{H}_{F \sigma}-\mu_{\sigma} \right) \hat{\psi}_{\sigma} + h
\hat{\psi}_{\sigma'}^{\dagger} \hat{\psi}_{\sigma'}
\hat{\psi}_{\sigma} + \sum_{\sigma'=\uparrow, \downarrow} \int
d^3r' \int_0^t d\tau W_{\sigma \sigma'}({\bf r},{\bf r}',\tau)
\phi_B({\bf r})\phi_B({\bf r}') \delta {\hat n}_{\sigma '}({\bf
r}', t-\tau) \hat{\psi}_{\sigma} + \hat{\Gamma}_{\sigma}
\hat{\psi}_{\sigma}. \label{fe}
\end{equation}
\end{widetext}
Here we have defined
\begin{eqnarray}
W_{\sigma \sigma'}({\bf r},{\bf r}',\tau)=\left(\frac{1}{i\hbar}
\right) \left[\Delta_{\sigma \sigma'}({\bf r},{\bf
r}',\tau)-\Delta_{\sigma\sigma'}^*({\bf r},{\bf r}',\tau)\right],
\nonumber
\end{eqnarray}
\begin{eqnarray}
\Delta_{\sigma \sigma'}= \frac{f_{\sigma} f_{\sigma'}}{V}
\sum_{\bf k} \sqrt{\frac{\epsilon_{\bf k}}{\epsilon_{\bf
k}+2gn_B}} e^{-iE_{\bf k} \tau/\hbar +i {\bf k} \cdot ({\bf
r}-{\bf r}')} , \nonumber
\end{eqnarray}
with $\epsilon_{\bf k}= \hbar^2 k^2/2m_B$, and $E_{\bf
k}=\sqrt{\epsilon_{\bf k}(\epsilon_{\bf k}+2gn_B)}$.

The third term in Eq.~(\ref{fe}) is the nonlinear optics analogue
of a fifth-order nonlinearity, involving two condensate fields,
the fermion density fluctuations which are quadratic in the
fermion fields, and the fermion field. Since these five fields can
mix to produce a sixth, the physical process involved is six-wave
mixing between the boson and fermion fields. Finally, the last
term in Eq.~(\ref{fe}), $\Gamma_{\sigma}=f_{\sigma} \phi_B({\bf
r})[ \hat{\xi}^{(0)}({\bf r},t)+ h.c.]$, is a stochastic potential
whose physical origin are the density fluctuations of the vacuum
state of the Bose quasi-particles, and has the two-point
correlation function $\langle \Gamma_{\sigma}({\bf r},t)
\Gamma_{\sigma'}({\bf r'},t') \rangle = n_B \Delta_{\sigma
\sigma'}({\bf r},{\bf r}',t-t')$.

To proceed we next assume that the fermion beam propagates through
the condensate at velocity $v$ less than the condensate sound
velocity $c=\hbar \sqrt{4\pi n_B a_B} /m_B$. Viewing the beam as
an impurity traversing the condensate, the condition $v<c$ implies
that it will not create incoherent phonon excitations that persist
in the condensate after the beam has passed \cite{TimmCote98}.
Physically, in this limit the fermion beam is accompanied as it
propagates by a virtual phonon cloud that mediates an effective
{\it instantaneous} interaction between the fermions. Furthermore,
since the phonon excitations are virtual the effects of the
stochastic potential $\Gamma_{\sigma}({\bf r},t)$ may safely be
neglected. Thus, we may neglect time retardation in the collision
term and the stochastic potential in the fermionic field
Eq.~(\ref{fe}). It then reduces to
\begin{widetext}
\begin{equation}
i\hbar \frac{\partial \hat{\psi}_{\sigma}({\bf r},t)}{\partial t}
= \left( \hat{H}_{F \sigma}-\mu_{\sigma} \right)
\hat{\psi}_{\sigma} + h \hat{\psi}_{\sigma'}^{\dagger}
\hat{\psi}_{\sigma'} \hat{\psi}_{\sigma} + \sum_{\sigma'=\uparrow,
\downarrow} \int d^3r' U_{\sigma \sigma'}({\bf r},{\bf r}')
\delta{\hat n}_{\sigma '} ({\bf r}',t) \hat {\psi}_{\sigma}({\bf
r}',t). \label{fef}
\end{equation}
\end{widetext}
where for the case of a homogeneous condensate $U_{\sigma
\sigma'}({\bf r},{\bf r}')$ is the Yukawa potential
\cite{stoof,smith}
\begin{equation}
U_{\sigma \sigma'}({\bf r},{\bf r}') = -\left (\frac{m_B
f_{\sigma} f_{\sigma'} n_B}{\pi \hbar^2}\right )
\frac{e^{-\sqrt{2}|{\bf r}-{\bf r}'|/l_h}}{|{\bf r}-{\bf r}'|}
\label{potential}
\end{equation}
and $l_h=\hbar/\sqrt{2m_B g n_B}$ is the healing length of the
Bose condensate. The minus sign in Eq.~(\ref{potential}) indicates
that the Fermi-Bose coupling induces an {\em attractive} effective
force between fermionic atoms. In the language of nonlinear optics
the condensate acts as a ``nonlinear crystal'' for the fermions
and the induced effective fermion-fermion interaction $U_{\sigma
\sigma'}({\bf r},{\bf r}',t)$ plays the role of a spatially
nonlocal Kerr nonlinearity for the fermionic field, leading to
familiar effects such as self-focusing. However, since fermionic
fields are intrinsically multimode and are not amenable to a
mean-field description \cite{note1} one cannot expect simple,
few-mode nonlinear effects such as occur in optics or bosonic atom
optics, but rather multimode coupling and many-particle quantum
correlations. Here we consider the particular case of formation of
two-fermion bound states, a problem that presents a close analogy
to the quantum propagation of a two-photon light beam in a
self-focusing medium \cite{chiao}.

\begin{figure}
\includegraphics*[width=0.95\columnwidth,
height=0.6\columnwidth]{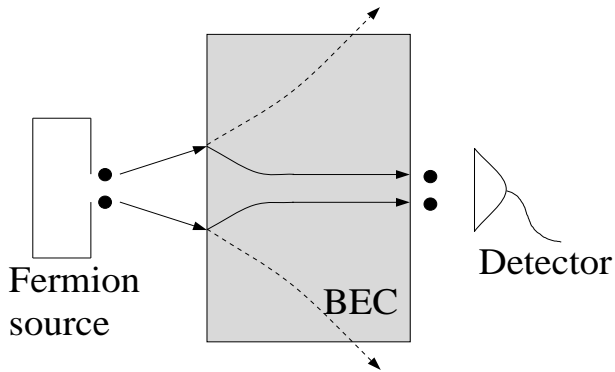} \caption{Schematic diagram for
the formation of the two-atom bound state in a fermionic beam
propagating through a Bose-Einstein condensate} \label{fig1}
\end{figure}

Figure 1 shows a potential nonlinear atom optics scheme to realize
the bound states. We assume that an ultracold source of fermionic
atoms generates a beam containing simultaneously two atoms. First
we consider a spin-polarized beam described by the two-atom
quantum state, $ |\Phi(t)\rangle =\frac{1}{\sqrt{2}} \int d^3r_1
\int d^3r_2 f({\bf r}_1,{\bf r}_2,t) \hat{\psi}^{\dagger}({\bf
r}_1) \hat{\psi}^{\dagger}({\bf r}_2)|0\rangle \label{spbeam}$,
that then propagates at velocity $v$ into a condensate. The
function $f({\bf r}_1,{\bf r}_2,t)$ gives the probability
amplitude to find one atom at ${\bf r}_1$ and the other at ${\bf
r}_2$, and is antisymmetric under atomic exchange. (We ignore the
spin label for simplicity of notation.) The time evolution of
$|\Phi(t)\rangle$ is determined by the Schr\"odinger equation
\begin{equation}
i\hbar \frac{\partial |\Phi\rangle}{\partial t}=H_{\rm eff}
|\Phi\rangle \label{seq}
\end{equation}
where the total effective Hamiltonian, constructed from
Eq.(\ref{fef}), is
\begin{eqnarray}
&&H_{\rm eff} = \int d^3r \hat{\psi}^{\dagger}({\bf
r})\left(H_F-\mu_F \right)\hat{\psi}({\bf r})\nonumber \\
&+& \frac{1}{2} \int\!\!\int d^3r d^3r' U({\bf r},{\bf r}')
\hat{\psi}^{\dagger}({\bf r})\delta{\hat n}({\bf r}')
\hat{\psi}({\bf r}). \label{heff}
\end{eqnarray}
This yields the equation of motion for $f({\bf r}_1,{\bf r}_2,t)$
\begin{widetext}
\begin{equation}
i\hbar \frac{\partial f({\bf r}_1,{\bf r}_2,t)}{\partial t} =
\left(H_F({\bf r}_1)+H_F({\bf r}_2) + U({\bf r}_1,{\bf r}_2) + 2
\int \!\! \int d^3r' d^3r'' \left[U({\bf r}_1,{\bf r}')+U({\bf
r}_2,{\bf r}')\right]|f({\bf r}',{\bf r}'',t)|^2 \right)f({\bf
r}_1,{\bf r}_2,t), \label{f12}
\end{equation}
\end{widetext}
which clearly shows the role of the effective Kerr medium on the
fermionic field. To determine the condition for formation of the
two-atom bound states of Eq. (\ref{f12}) we transform into the
center of mass coordinates ${\bf R}=({\bf r}_1+{\bf r}_2)/2$ and
${\bf r}={\bf r}_1-{\bf r}_2$, and make the ansatz $f({\bf
r}_1,{\bf r}_2,t)=e^{i {\bf K} \cdot {\bf R}-iEt/\hbar}W({\bf
r})$, which leads finally to the eigenvalue problem for the
relative motion of two particles in a Yukawa potential
\begin{equation}
\left(-\frac{\hbar^2 \nabla_r^2}{2 \mu} -\frac{U_0}{r}
e^{-\sqrt{2} r/l_h}\right)W({\bf r})=E_{r}W({\bf r}), \label{rmeq}
\end{equation}
where $U_0 \equiv f^2 m_B n_B/(\pi \hbar^2)$ is the potential
strength, $\mu=m_F/2$ the reduced mass, and $E_{r}$ the energy of
relative motion. From Eq.(\ref{rmeq}), one can see that the
spatial width of the energy eigenstates is dictated by the length
scale $l_0=\hbar^2/(\mu U_0)$. The energy eigenvalues of a Yukawa
potential have been extensively investigatetd. An accurate formula
\cite{green} has been provided for the number of bound states for
a state with total angular momentum quantum number $l$,
\begin{equation}
\nu=(\sqrt{Z}-\sqrt{Z_l})S_l + 1, \label{boundnumber}
\end{equation}
where $Z=l_h/(\sqrt{2} l_0), Z_l=0.8399(1+2.7359 l+1.6242 l^2)$
and $S_l=1.1335(1+0.0191 l-0.001684 l^2)$. Considering the
requirement of symmetry of the spatial wave function for the
spin-polarized beam, the angular momentum must be $l=1,3,5...$.
The condition that there exists at least one bound state is $Z
\geq Z_1$. This determines a spatial range $l_0 \leq l_{b} \equiv
l_h /6.366$ for two-atom bound state in a Bose condensate, and
meanwhile enforces a requirement of sufficient potential strength
$U_0 \geq U_b \equiv 6.366 \hbar^2/(\mu l_h)$ for binding.

Possible candidates to observe two-fermion bound states include
combinations of alkali atoms such as $^{6}$Li-$^{7}$Li,
$^{6}$Li-$^{23}$Na, and $^{40}$K-$^{39}$K. The $^{40}$K-$^{39}$K
combination appears to be the most promising, due to its very
large boson-fermion scattering length $a_{BF} \sim$ 52.9 nm and
its very small boson-boson scattering length $a_B \sim$ 0.26 nm.
In this case, the realization of bound fermionic states requires a
condensate density of $n_B \sim 10^{13}$ cm$^{-3}$, a value
achievable with current techniques.

The basic signature of the formation of a two-fermion bound state
in the condensate is that it is immune to wave packet spreading
and can therefore be collected at a localized detector placed at
the output of the condensate as illustrated in Fig. 1. In
contrast, if the condensate is absent, and the time of flight of
the fermions is chosen long enough, the initial fermion packet
will spread out such as to negligibly overlap the detector. The
signature of the formation of bound states in this nonlinear atom
optics arrangement is therefore an enhanced detection of atoms
when the condensate is in place in contrast to when it is absent.
This remains true even though the initial fermion wave packet does
not exactly match the bound state, as long as it has some
reasonable projection onto the bound state wave function.

We remark that our results are readily generalized to fermionic
atomic beam with two internal spin states. In particular, if one
initially prepares the beam in a spin-singlet state, the two-atom
quantum state will have the form $|\Phi\rangle=\frac{1}{2}
\int\!\!\int d^3r_1 d^3r_2 f({\bf r}_1,{\bf r}_2,t)
(\psi_{\uparrow}^{\dagger}({\bf r}_1)
\psi_{\downarrow}^{\dagger}({\bf
r}_2)-\psi_{\downarrow}^{\dagger}({\bf r}_1)
\psi_{\uparrow}^{\dagger}({\bf r}_2) )|0\rangle$. Then in addition
to the attractive Yukawa potential the two fermions also
experience an $s$-wave interaction, since they are in different
spin states. Nonetheless two-fermion bound states can still be
formed under suitable combination of these potentials. In
particular, the formation of the spatial bound states in the
spin-singlet case may be useful to create the spatially entangled
fermionic atomic beam.

In conclusion, we have derived a nonlinear Schr\"odinger equation
for a fermionic beam in a condensate, and shown that the
condensate acts as an effective nonlinear medium for the fermions,
leading to formation of a two-atom bound state closely analogous
to the two-photon bound state in self-focusing media \cite{chiao}.
A clear signature of the bound state in a nonlinear atom optics
experiment has also been proposed. Physically, the understanding
of fermionic bound states may be important for the manipulation of
the quantum statistical properties of fermionic atomic beams, e.g.
changes from antibunched and bunched beams, dynamic Cooper
pairing, and potentially the formation of quantum solitons in
ultracold fermionic atomic beams.

We thank Eddy Timmerman for numerous discussions. This work is
supported in part by the US Office of Naval Research under
Contract No. 14-91-J1205 and No. N00014-99-1-0806, by the National
Science Foundation under Grant No. PHY98-01099, by the US Army
Research Office, and by the Joint Services Optics Program.


\begin{references}
\bibitem{note1} An exception is that of fermionic atoms that form
Cooper pairs after undergoing a BCS phase transition. The Cooper
pairs can be described using the mean-field Ginzburg-Landau
theory.
\end{references}
\end{document}